\newif\ifportable	
\portabletrue		

\newif\iffigs
\figsfalse		

\ifportable
	\makeatletter
	\@ifundefined{documentclass}{	
		\iffigs
		\documentstyle[oldlfonts,epsfig]{article}
		\else
		\documentstyle[oldlfonts]{article}
		\fi
		\def\emph#1{{\em #1\/}}	
		}{
		\documentclass[a4paper,twocolumn]{article}	
		\iffigs\usepackage{epsfig}\fi
		\usepackage{times}
		}
	\iffigs\psfigdriver{oztex}\fi
	\makeatother
\else				
	\documentstyle{l-aa}
	\def\emph#1{{\em #1\/}}
\fi

\makeatletter
\let\@internalcite\cite
\def\cite{\@ifstar{\citeyear}{\citefull}}
\def\citefull{\def\astroncite##1##2{##1, ##2}\@internalcite}
\def\citeyear{\def\astroncite##1##2{##2}\@internalcite}

\def\@citex[#1]#2{\if@filesw\immediate\write\@auxout{\string\citation{#2}}\fi
  \def\@citea{}\@cite{\@for\@citeb:=#2\do
    {\@citea\def\@citea{; }\@ifundefined
       {b@\@citeb}{{\bf ?}\@warning
       {Citation `\@citeb' on page \thepage \space undefined}}%
{\csname b@\@citeb\endcsname}}}{#1}}

\def\@cite#1#2{(#1\if@tempswa , #2\fi)}
\def\@biblabel#1{}
\makeatother

\title{Particle acceleration and entropy considerations%
\ifportable\thanks{To appear in A\&A.  Preprint no. Astro-95/02, and
	astro-ph/9501083.}\fi
}
\ifportable
	\author{John C Brown$^\dag$, Gregory Beekman$^\dag$, Norman Gray\thanks
		{Dept.\ of Physics and Astronomy, University of Glasgow,
		Glasgow G12 8QQ, U.K.} \\
	Alexander L MacKinnon\thanks{Dept.\ of Adult and Continuing Education,
		University of Glasgow, Glasgow G12 8QQ, UK}}
	\date{12 January 1995}
\else
	\author{John C Brown\inst{1}\and Gregory Beekman\inst{1}\and
	Norman Gray\inst{1}\and Alexander L MacKinnon\inst{2}}
	\institute{Dept.\ of Physics and Astronomy, University of Glasgow,
		Glasgow G12 8QQ, U.K.\and
		Dept.\ of Adult and Continuing Education, University of Glasgow,
		Glasgow G12 8QQ, UK}
	\offprints{John C Brown}
	\date{Received date; accepted 12 January 1995}
	\thesaurus{09(02.01.1; 06.06.3; 09.03.2)}
\fi

\ifportable
\fi

\newcommand{\E}{E}		
\renewcommand{\d}{{\rm d}}
\newcommand{\be}[1]{\begin{equation}\label{e:#1}}
\newcommand{\ee}{\end{equation}}
\newcommand{\bea}{\begin{eqnarray}}
\newcommand{\eea}{\end{eqnarray}}

\def\:#1#2{{\scriptstyle {#1\over #2}}}
\def\eqnref#1{(\ref{e:#1})}
\newcommand{\du}{\delta^3\Vec u}

\ifportable
\newcommand{\la}{\mathrel{\vcenter{\hbox{\ooalign{\raise 4.75pt
	\hbox{$<$}\crcr $\sim$}}}}}
\fi

\ifportable
\def\Vec#1{{\bf #1}}
\else
\let\Vec\vec			
\fi

\begin{document}
\maketitle

\begin{abstract}
Possible entropy constraints on particle acceleration spectra are
discussed.

Solar flare models invoke a variety of initial distributions of the
primary energy release over the particles of the flare plasma -- ie., the
partition of the energy between thermal and nonthermal components.  It is
suggested that, while this partition can take any value as far as energy is
concerned, the entropy of a particle distribution may provide a useful
measure of the likelihood of its being produced for a prescribed total
energy.

The Gibbs' entropy is calculated for several nonthermal isotropic
distribution functions~$f$, for a single particle species, and compared
with that of a Maxwellian, all distributions having the same total number
and energy of particles.  Speculations are made on the relevance of
some of the results to the cosmic ray power-law spectrum, on their relation
to the observed frequency distribution of nonthermal flare hard X-ray
spectrum parameters and on the additional energy release required to
achieve lower entropy~$f$s.

\ifportable
\else
\fi

\end{abstract}

\section{Introduction}

Models in which energetic particles are invoked as a major component of
primary energy release in the impulsive phase of solar flares have been
both popular and controversial for decades (see eg, reviews
\cite{brown91,simnett91}).  Despite ever improving data and theoretical
modelling, however, it remains unresolved how important such particles
are in flare energy transport.  Brown and
Smith~\cite*{brownsmith80} suggested that the lower entropy of accelerated
particles compared to a pure Maxwellian distribution might place a
theoretical thermodynamic constraint on the fraction of magnetic energy
release likely to go into such acceleration.  Brown~\cite*{brown93} has
raised this issue again and illustrated the point with a simple example
of the thermal/nonthermal entropy difference.

In this paper we put this question on a more quantitative basis by examining
the entropy of more general particle distribution functions.  Essentially
the idea is that, solely on energy conservation grounds (first law of
thermodynamics), deposition of energy~$\E$ among~$N$ cold particles could
equally well result in (for example) a Maxwellian of temperature~$2\E/3Nk$,
a single particle of energy~$\E$ with the rest cold, or a delta function
distribution of all the particles at particle energy~$\E/N$.  In a
statistical sense, however, it is obvious that the last two are much less
probable than the first -- ie., they have much lower entropy. (Evolution of
a plasma distribution function~$f$ from a cold Maxwellian to a nonthermal
form with high mean energy does of course involve
a particle entropy increase and so is not
precluded by the second law of thermodynamics.  However, while not
\emph{impossible} such an outcome is statistically \emph{much less probable}
than production of a hot Maxwellian~$f$ since it has lower entropy).
Quantifying how much lower the entropy (probability) is should help
constrain, in a statistical sense, the likely forms of particle
distributions produced by prescribed energy release conditions or
conversely constrain the energy release conditions required to yield a
prescribed particle distribution.

We recognise at the outset that the present calculations are exploratory
and rest on simplifying assumptions which need further study.  In
particular we assume that entropy considerations are relevant -- this
raises issues regarding equilibrium and closure of the system, for example,
which have been addressed in broad terms by Tolman
\cite*[pp.~560-564]{tolman} -- but we feel that this will be so provided we
deal only with newly-accelerated particles \emph{outside} the (presumably
small) accelerating volume.  Secondly, we compare here only the entropies
of alternative forms of the \emph{particle} distribution function which
could be generated by deposition of energy~$\E$ among~$N$ particles in a
prescribed volume~$V$.  We neglect, for the moment, entropy contributions
from other system degrees of freedom and, in particular, from plasma waves,
and from the changing magnetic field presumed responsible for generating
the flare particle distribution.  The possible importance of including
plasma waves as additional degrees of freedom in considering the evaluation
of plasma entropy during particle acceleration has been discussed
in~\cite{grognard}.  Our neglect of magnetic field entropy is based on the
heuristic argument that the field is associated solely with an ordered zero
entropy flow component of the particles -- the Gibbs entropy $\sim
\int\!\!\int\!\!\int f\ln f \,\d^3\Vec v$ of any particle distribution
function~$f(\Vec v)$ is unchanged by adding a systematic drift~$\Vec v_0$
(ie., $\Vec v \to \Vec v+\Vec v_0$).


\section{Definitions}
\label{s:assumptions}

In this paper we will deal solely with the Gibbs entropy~$S$ of the 3-D
(velocity space) distribution function~$f(\Vec  r,\Vec  v)$ for a single
species of nonrelativistic particle of mass~$m$, defined by
\be{entropydef}
	S = -k \int_{\Vec v} \int_{\Vec r} f \ln f \,\d^3\Vec r\, \d^3\Vec v
\ee
where~$k$ is Boltzmann's constant.
For simplicity, we consider only distributions
in which the velocity~$\Vec v$ distribution is separable from the
configuration space~$\Vec r$ dependence.
With this restriction, we may write
\be{fng}
	f(\Vec r,\Vec v) = n(\Vec r) G(\Vec v)
\ee
where~$n(\Vec r)$ is the total particle space density at~$\Vec  r$ and
\be{Gdv}
	\int_{\Vec v} G(\Vec v)\,\d^3 \Vec v = 1
\ee

Because~\eqnref{entropydef} is not linear in~$f$, the absolute value
of~$S$ depends on the velocity and coordinate units used.  Here, however,
we will only be concerned with entropy differences between
different~$G(\Vec v)$. If we take some characteristic speed~$v_0$ as
velocity unit and write~$\Vec v = v_0 \Vec u$ and
$G(\Vec v) \to g(\Vec u)/v_0^3$
then~\eqnref{entropydef} and~\eqnref{Gdv} yield
\begin{eqnarray}
	S & = & -k \int_{\Vec u} \int_{\Vec r} n(\Vec r) g(\Vec u)
		\ln [n(\Vec r) g(\Vec u)/v_0^3] \,\d^3\Vec r \d^3\Vec u
		\nonumber\\
	  & = &   - k \int_{\Vec r} n(\Vec r) \ln n(\Vec r) \,\d^3\Vec r
	  	\int_{\Vec u} g(\Vec u) \,\d^3\Vec u
	  	\nonumber\\
	  &   & \quad{}- k \int_{\Vec r} n(\Vec r) \,\d^3
	  	\Vec r \int_{\Vec u} g(\Vec u) \ln g(\Vec u) \,\d^3\Vec u
	  	\nonumber\\
	  &   & \quad{}+ k \ln v_0^3 \int_{\Vec r} n(\Vec r) \,\d^3r
	  	\int_{\Vec u} g(\Vec u) \,\d^3\Vec u
	  	\nonumber\\
	  & = & -kN_0 \int_{\Vec u} g(\Vec u) \ln g(\Vec u) \,\d^3\Vec u
	  	\nonumber\\&&\quad{} 
	  	+ kN_0 \left\{ \ln v_0^3 - {1\over N_0}
	  	\int_{\Vec r} n(\Vec r) \ln n(\Vec r)
	  	\,\d^3\Vec r \right\}
	  	\label{e:Sint}
\end{eqnarray}
where
\be{N0int}
	N_0 = \int_{\Vec r} n(\Vec r) \,\d^3\Vec r
\ee
is the total number of particles.  If we consider different distribution
functions~$f$ with the same \emph{spatial} distribution~$n(\Vec r)$ and total
volume (ie., same~$N_0$) and use the same velocity unit~$v_0$ throughout
then we can measure the \emph{entropy differences} by the `scaled entropy'
\be{sigmadeforig}
	\Sigma = {1\over k N_0} \left( S - k N_0 \left\{ \ln v_0^3
		- {1\over N_0} \int_{\Vec r} n(\Vec r) \ln n(\Vec r) \,\d^3
		\Vec r\right\} \right)
\ee
so that by~\eqnref{N0int},~$\Sigma$ can be expressed solely in terms
of~$g(\Vec u)$, via
\be{sigmadef}
	\Sigma\{g\} = - \int_{\Vec u} g(\Vec u) \ln g(\Vec u)\,\d^3\Vec u
\ee
and we note, in terms of~$g$, normalisation~\eqnref{Gdv} is
\be{gdv}
	\int_{\Vec u} g(\Vec u) \,\d^3\Vec u = 1.
\ee

We do not consider further here the spatial structure contribution to the
entropy (last term in \eqnref{sigmadeforig}) -- i.e., we only consider
entropies of different $g(u)$ but the same $n(\Vec r)$.Also, throughout the
rest of this paper we consider only isotropic~$f(\Vec v)$ so that
$\int_{\Vec u}\d^3\Vec u \to \int_0^\infty 4\pi u^2\,\d u$.  We will be
comparing the relative entropies of various distributions~$g$ with that of
a pure Maxwellian of the same~$N_0$ and of the same total energy~$\E$ (ie.,
we are comparing identical plasmas with the same total energy deposited in
them).  It is therefore convenient to use as velocity unit~$v_0$, the
thermal speed in the Maxwellian plasma
\be{v0}
	v_0 = (2kT_0 / m)^{1/2}
\ee
where~$T_0 = 2\E/3N_0 k$ is the temperature of the plasma when all
of~$\E$ is deposited as heat in a pure Maxwellian.  With this choice
of~$v_0$ and with~$g$ isotropic the normalisation condition~\eqnref{gdv}
ensuring the same number of particles is that~$g(\Vec u)$ should satisfy
\be{gu2du}
	\int_0^\infty g(u) u^2 \,\d u = {1\over 4\pi}
\ee
while the constraint ensuring the same total energy becomes
\be{gu4du}
	\int_0^\infty g(u) u^4 \,\d u = {3\over 8\pi}.
\ee

\section{Scaled entropies of specific distribution functions}

\subsection{Pure Maxwellian $g_M(u)$}
\label{ss:Maxwellian}

Expressed in terms of $g$, a pure Maxwellian satisfying
constraints~\eqnref{gu2du} and~\eqnref{gu4du} is
\be{gM}
	g_M(u) = {e^{-u^2} \over \pi^{3/2}}
\ee
which, in~\eqnref{sigmadef}, integrates straightforwardly to give the
scaled entropy
\be{sigmaM}
	\Sigma_M = \frac32 (1 + \ln\pi).
\ee

\subsection{Maxwellian with bump in tail $g_{BIT}(u)$}
\label{ss:BIT}

One of the forms of distribution function with an accelerated
component is a Maxwellian with a `bump in tail'.  Here we consider only a
simple case to allow analytic treatment, namely a bump in tail centred on
speed~$u_1$, and of narrow width~$\Delta u \ll 1$ over which the bump
contribution to~$g$ is a constant added to the Maxwellian component which
describes the rest of the plasma (cf Brown~\cite*{brown93}).
Because of energy constraint~\eqnref{gu4du}, the temperature of this
Maxwellian component is reduced by a factor~$\tau<1$ relative to the pure
Maxwellian of case~\ref{ss:Maxwellian}, by the energy resident in the bump.
If the fraction
of the particles by number in the bump and in the Maxwellian are
respectively~$\phi$ and~$(1-\phi)$ (to satisfy the number
constraint~\eqnref{gu2du}) then
\be{gBIT}
	g_{BIT}(u) = {1-\phi \over (\pi\tau)^{3/2}}
			 e^{-u^2/\tau} +
			 \left\{
			 \begin{array}{cl}
			 	\displaystyle {\phi \over 4\pi u_1^2 \Delta u}
			 		& u\hbox{ in }\Delta u  \\
			 	0 & \hbox{otherwise}
			 \end{array}
			 \right.
\ee
with
\be{taudef}
	\tau = {1 - \:23 \phi u_1^2 \over 1-\phi}
\ee
and we note that the condition $u_1^2 \le 3/2\phi$ has to be met
since~$\tau\ge0$, equality holding when the Maxwellian component
of~$(1-\phi)N_0$ remains cold.

Substituting~\eqnref{gBIT} and~\eqnref{taudef} in~\eqnref{sigmadef} and
approximating the integrals on the assumption that~$\Delta u\ll1$ and
$\phi/(4\pi u_1^2\Delta u) \gg {1-\phi e^{-u_1^2/\tau} /
(\pi\tau)^{3/2}}$ (ie, the bump is locally `large') we obtain for the
scaled entropy
\bea\label{e:sigmaBIT}
	\rlap{$\displaystyle \Sigma_{BIT}(\phi,u_1,\Delta u) ={}$}\quad
		&&\nonumber\\&&
		- \phi \ln\left({\phi\over4\pi u_1^2\Delta u}\right)
		+ \frac32 (1-\phi)			\nonumber\\
		&&{}\times\left[
				1 + \ln\pi - \frac53\ln(1-\phi)
					+ \ln \left(1-\frac23 \phi u_1^2 \right)
				\right].
\eea

\subsection{Pure power law $g_{PL}(u)$}
\label{ss:PL}

A commonly occurring form of accelerated particle spectrum in solar flares
and elsewhere in astrophysics is the power law~$v^{-\alpha}$.  Such~$f(v)$
diverges unless flattened or truncated at small~$v$, so we first consider a
truncated pure power law comprising \emph{all} the plasma particles and
again with the same total energy as the pure
Maxwellian~\ref{ss:Maxwellian}.  With number constraint~\eqnref{gu2du},
this has the form
\be{gPL}
	g_{PL} = \cases{0 & $u<u_1$\cr
		\displaystyle {\alpha-3\over 4\pi u_1^3}
		\left({u\over u_1}\right)^{-\alpha} & $u\ge u_1$\cr}
\ee
where the dimensionless low energy cut-off speed~$u_1$ has to be related
to~$\alpha$ to satisfy total energy constraint~\eqnref{gu4du}, \emph{viz.}
\be{u1PL}
	u_1 = \left({3\over2} {\alpha-5\over\alpha-3}\right)^{1/2}
\ee
and we note that~$\alpha>5$ is required for finite total energy.  We note
that the relation between~$\alpha$ and the spectral index~$\delta$ normally
used (in solar flare physics) to describe the non-relativistic particle
flux spectrum differential in kinetic energy \cite{brown71}
is~$\alpha=2\delta+1$.

Substitution of~\eqnref{gPL} and~\eqnref{u1PL} in~\eqnref{sigmadef} leads to
the scaled entropy
\be{sigmaPL}
	\Sigma_{PL}(\delta) = {\delta+\:12\over\delta-1}
				+ \:12\ln\left[{27\pi^2\over2}
					{(\delta-2)^3\over(\delta-1)^5}\right]
\ee

\subsection{Maxwellian with a power law tail}
\label{ss:MPL}

The conventional wisdom regarding the behaviour of the real distribution
funtion for the bulk of the electrons in a solar flare is that it is
roughly Maxwellian at low velocities (where collisions dominate) but of
power law (or other nonthermal form) at high velocities (where mean free
paths become long and runaway may occur).  Such a particle model spectrum
finds some support in recent inversions of high resolution bremsstrahlung
hard x-ray spectra \cite{johns92,thompson92,piana93} though these
inferences are ambiguous because of the effects of noise \cite{craig86}
and of averaging over an inhomogeneous source \cite{brown71}.  It is
therefore instructive to consider such a distribution from the viewpoint of
entropy.  A convenient functional form which exhibits these asymptotic
behaviours at low and high~$v$, and which circumvents most of the analytic
messiness associated with a piecewise description
is~$f(v)\sim(1+v^2/v_1^2)^{-\alpha/2}$~\cite{paterson93}.  The high
velocity spectral index is~$\alpha$, as in~\ref{ss:PL}, and the
temperature~$T$ defining the shape at small~$v$ is fixed by~$\alpha$
and~$v_1$ such that~$\:12mv_1^2=\alpha kT$.  Similarly to
cases~\ref{ss:BIT} and~\ref{ss:PL}, $\alpha$ and~$v_1$ have to be
interrelated to satisfy constraints~\eqnref{gu2du} and~\eqnref{gu4du},
namely $u_1=v_1/v_0=\sqrt{(\alpha-5)/2}$.  Taking these constraints into
account, the resulting~$g(u)$ is
\be{gMPL}
	g_{MPL}(u) = \left[{2\over \pi(\alpha-5)}\right]^{3/2}
		 {\Gamma\left({\alpha\over2}\right)
		 	\over\Gamma\left({\alpha-3\over2}\right)}
			\left(1+{2\over\alpha-5}u^2\right)^{-\alpha/2}
\ee
where~$\Gamma(x)$ is the gamma function.

Substitution of~\eqnref{gMPL} in~\eqnref{sigmadef} gives, for this case
(with~$\alpha=2\delta+1$ as in Sect.~\ref{ss:PL})
\bea\label{e:sigmaMPL}
	\Sigma_{MPL}(\delta) &=& -\ln \left[
			{\Gamma(\delta+\:12)\over
			\pi^{3/2}(\delta-2)^{3/2}\Gamma(\delta-1)}\right]
			\nonumber\\&&\quad{} 
			- (\delta+\:12)[\psi(\delta-1) - \psi(\delta+\:12)]
\eea
where the psi function is
\be{psidef}
	\psi(x) \equiv {\d\over\d x}[\ln\Gamma(x)].
\ee

\section{Entropy comparison of nonthermal distributions with a pure Maxwellian}

We now compare the entropies of nonthermal
distributions \eqnref{sigmaBIT}, \eqnref{sigmaPL}, and \eqnref{sigmaMPL}
with that of the pure Maxwellian~\eqnref{sigmaM}.

\subsection{Bump-in-tail}
\label{ss:compareBIT}

This case was already briefly discussed by Brown~\cite*{brown93}.  Its
most important features are:
\begin{enumerate}
	\item  As~$\phi\to0$, the pure Maxwellian entropy~\eqnref{sigmaM} is
	recovered.

\item For \emph{any}~$\phi$, $u_1$ (ie, no matter how few particles are
accelerated or to how low a speed), the relative entropy~${}\to-\infty$
as~$\Delta u\to0$ -- ie, as a finite number of particles in~$f(u)$ is
crowded into an arbitrarily narrow velocity range.  Considering the
relative entropy as the negative of the log of the probability $P$,
this quantifies the qualitative improbability
statement in Sect.~1 concerning the relative improbability of nonthermal
distributions.  We note, however, that the divergence of $\Sigma$ to
$-\infty$ is an artefact of taking the approximate continuum function
expression (1) for entropy to a point where the smoothness assumption made
breaks down (see Appendix).

	\item  No matter how small the fraction~$\phi$ of particles accelerated
	is, as~$u_1^2 \to \:32\phi$, $\Sigma_{BIT}\to -\infty$ because the
	fraction of the total energy in fast particles~${}\to1$ at that point,
	which is arbitrarily improbable. The same remark as in~2 applies.
\end{enumerate}

\subsection{Pure power law}
\label{ss:comparePL}

In figure~\ref{f:1}a,
\ifportable
\begin{figure}
	\iffigs
		\epsfig{file=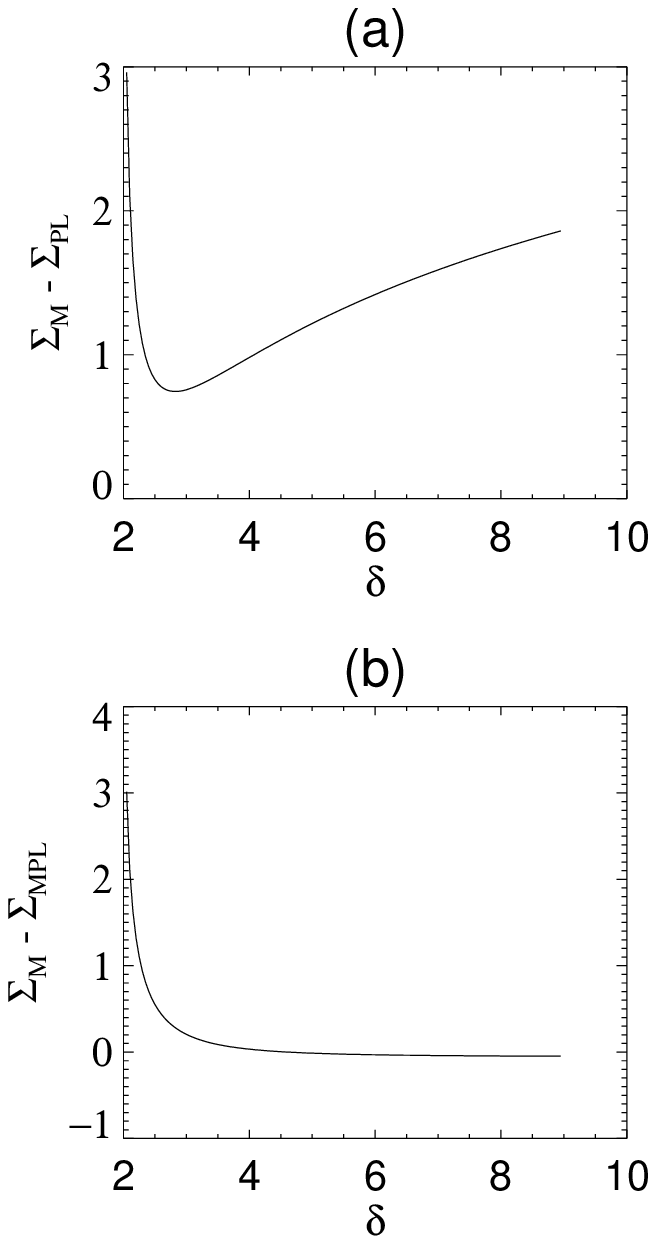,height=226pt}
	\else
		\vbox to 226pt{\hrule
			\vfill\hbox{Figure 1 is in file fig1.ps}\vfill\hrule}
	\fi
	\caption{\protect\label{f:1}Scaled entropy
	$\Delta\Sigma\equiv\Sigma_M-\Sigma$,
	relative to
	the Maxwellian for (a) a pure truncated power law and
	(b) a Maxwellian with a power-law tail.}
\end{figure}
\else
\begin{figure}
	\picplace{130mm}
	\caption[ ]{Scaled entropy $\Delta\Sigma\equiv\Sigma_M-\Sigma$,
	relative to
	the Maxwellian for (a) a pure truncated power law and
	(b) a Maxwellian with a power-law tail.}
	\protect\label{f:1}
\end{figure}
\fi
we show the
quantity~$\Delta\Sigma_{PL}(\delta)=\Sigma_M-\Sigma_{PL}$ as a function
of~$\delta$ for the pure power-law case.  The key features of this curve are
\begin{enumerate}
	\item  There is a minimum in~$\Delta\Sigma_{PL}$ which occurs
	at~$\delta=\delta_0\equiv(3+\sqrt7)/2\approx2.82$.  This is the value
	of~$\delta$ at which a sharply truncated power law most closely
	approaches the entropy of the Maxwellian.

	\item As~$\delta\to\infty$, $\Sigma_{PL}\to-\infty$ and
	$\Delta\Sigma_{PL}\to+\infty$.  This is because as~$\delta\to\infty$,
	more and more of the particles are concentrated at a single
	velocity~$u_1=\sqrt{3/2}$ (by~\eqnref{u1PL})
	and the distribution is more
	and more like a delta function, with associated low probability
	(cf.,~\ref{ss:compareBIT}).
	The same limitation applies here as in point~2
	of Sect.~\ref{ss:compareBIT}.

	\item  As~$\delta\to2$, $\Delta\Sigma_{PL}\to+\infty$ logarithmically.
	This is because the distribution function then becomes very flat and
	wide with a decreasing probability of the increasing number of particles
	needed at~$u\to\infty$ to satisfy the total energy
	constraint~\eqnref{gu4du}.
\end{enumerate}

Another way to express the entropy comparison between~$g_M$ and~$g_{PL}$ is
to consider the temperature~$T_\delta$ (or total energy) which a pure
Maxwellian would need to
have in order to have as low an entropy as~$g_{PL}$ with the same total
number~$N_0$ of particles.  The generalisation of~\eqnref{gM} to a Maxwellian
of temperature~$\tau=T/T_0$
is~$g_M(u,\tau)=\exp(-u^2/\tau)/(\pi\tau)^{3/2}$ and
of~\eqnref{sigmaM} is
\be{sigmaMtau}
	\Sigma_M(\tau) = \:32[1+\ln(\pi\tau)].
\ee

Equating~\eqnref{sigmaMtau} with~\eqnref{sigmaPL}
we can find the dimensionless temperature
{}~$\tau(\delta)=T_\delta/T_0$ required for the Maxwellian entropy to be
as low as the power law, \emph{viz.}
$\ln\tau_{PL}=2[\Sigma_{PL}-\Sigma_M]/3$, or
\be{tauPL}
\ln\tau_{PL}(\delta) = {4-\delta\over3(\delta-1)}
			+ \ln\left[{3\over(2\pi)^{1/3}}
			{\delta-2\over(\delta-1)^{5/3}}\right]
\ee
which also has a maximum at the value of~$\delta=\delta_0$.

The above results allow some quantitative interpretation, albeit somewhat
speculative.  First, if we consider transient energy release events, then
the entropies of distribution functions of different~$\delta$ should be
measures of the relative probabilities of these~$\delta$s being realised
\emph{if} all the energy went into a pure power-law.  In a large set of
such events these should reflect in some way the relative frequency of
occurrence of different~$\delta$s.  In particular, values of~$\delta$
near~$\delta_0$, where~$\Sigma_{PL}$ has a maximum, should be commonest
(This is certainly not the case in the observed frequency distribution of
flare~$\delta$ values inferred from hard X-ray bursts, but nor are flare
spectra those of pure power laws with no Maxwellian component! -- cf.\
Sect.~\ref{ss:compareMPL}).


Second, we note that the above power-law entropy analysis also applies to
the relativistic regime when $v$ is replaced by the momemtum $p$ and the
maximum entropy power-law spectrum again has index $\delta_0$.  This
$\delta_0$ is very close to the observed mean spectral index of cosmic ray
particles over a very wide energy range~\cite{longair} -- a fact pointed
out to us by Bell~\cite*{bell}.  That is, if an unspecified mechanism
operates to force a power-law form on the distribution function, under
conditions of constant total particle number and energy, then the highest
entropy (most likely) state is that with a power-law index close to the
observed cosmic-ray one.  This fact should be explored further.

\subsection{Maxwellian with power-law tail}
\label{ss:compareMPL}

In figure~\ref{f:1}b we show the variation with $\delta$
of~$\Delta\Sigma_{MPL}(\delta)\equiv\Sigma_M-\Sigma_{MPL}$ (equivalent
to~$\tau_\delta$) for this case, defined similarly to those
in~\ref{ss:comparePL}, based on equations~\eqnref{sigmaM}
and~\eqnref{sigmaMPL}, \emph{viz.}
\bea
	\Delta\Sigma_{MPL}(\delta) &=& \frac32 +
		\ln\left[ {\Gamma(\delta+\:12)\over
		(\delta-2)^{3/2}\Gamma(\delta-1)} \right]\nonumber\\
	&&\quad {}+ (\delta+\:12) [\psi(\delta-1) - \psi(\delta+\:12)]
						\label{e:deltasigmaMPL}\\
\eea

In this case, which is expected to be much closer to the solar flare
 situation than the pure
power law, there is no extremum in~$\Delta\Sigma_{MPL}$ which tends
to~$+\infty$ as~$\delta\to2$ and falls
as~$\delta$ increases. This is essentially because, for a prescribed
total~$N_0$ and~$\E$, the steep power law tail more and more closely
resembles the Maxwellian for large~$\delta$.  On the other hand
as~$\delta$ decreases, the entropy becomes smaller and smaller as
deviation from the Maxwellian increases.  If this result is interpreted
statistically in terms of the relative likelihood of a tail of
given~$\delta$ occurring, as compared to pure heating, it means that
smaller and smaller~$\delta$ should occur less and less frequently.

Observational inference of~$\delta$ in solar flares is achieved from the
spectral index~$\gamma$ of bremsstrahlung hard X-rays
(eg, Brown~\cite*{brown71}). In terms of the electron acceleration spectrum the
event integrated~$\gamma$ is related to~$\delta$ by~$\delta=\gamma+1$
(for collision dominated thick target energy losses) and by ~$\delta=\gamma-1$
for collisionally thin targets.  Kane~\cite*{kane} has reported
the statistics of observed occurrence of different~$\gamma$ values.  The
qualitative trend of these data is similar to that predicted from the
above entropy argument -- ie, a sharp decline of numbers of events of
small spectral index and a frequency distribution flattening off at
larger indices (though the data are instrumentally limited to
indices~$\gamma\la6$).

In view of the simplications made in the present
theoretical analysis (homogeneous source volume, neglect of waves and of
anisotropy) the trend of the theoretical and observed distributions
is remarkably similar, and worthy of further investigation.

\section{Conclusions}
\label{s:conclusions}

The lower entropy of a hot plasma with an accelerated component, as
compared to a pure Maxwellian of the same total particle number and energy,
quantifies the extent to which the hardest non-thermal components are
further from equilibrium and so should occur less frequently.  Calculation
of the entropy difference in the case of a Maxwellian with a power-law tail
of index~$\delta$ leads to a predicted frequency distribution of spectral
indices in qualitative agreement with the general trend of the distribution
inferred from the statistics of flare hard X-ray bursts.

All distributions except the Maxwellian are non-equilibrium ones, which
means that we cannot define a temperature for them and so, in turn, that
we cannot use our expression for the entropies of these distributions in any
non-trivial purely thermodynamical argument.  We can, however, make a
cautious statistical argument, as we have done in
Sects.~\ref{ss:comparePL} and~\ref{ss:compareMPL}, to the extent that,
given a process which produces a certain non-thermal particle
distribution, we expect some characteristics of that distribution to be
more likely than others.  We have no doubt that there are more, and more
detailed, arguments to be made using these results.

The close coincidence of the observed cosmic-ray spectral index with that
of a maximum entropy pure power-law is likewise tantalising.

\ifportable
\section*{Acknowledgements}
\else
\begin{acknowledgements}
\fi

GB gratefully acknowledges the support of a Royal Society of Edinburgh
Cormack Vacation Research Scholarship.  JCB, NG,  and ALM gratefully
acknowledge
the support of a PPARC grant, an EEC Contract, and of the Research Centre
for Theoretical Astrophysics, University of Sydney, where part of this
work was carried out.  The paper has benefited from discussions of the
entropy problem with numerous colleagues and in particular with D~B
Melrose, P~Robinson, A.M.Thompson, and A~R Bell.

\ifportable
\section*{Appendix: the limitations of Eq.\ \protect\eqnref{sigmadef}}
\else
\end{acknowledgements}
\appendix
\section{The limitations of Eq.\ \protect\eqnref{sigmadef}}
\fi
As pointed out in Sect.~\ref{ss:compareBIT}, the divergence of the
`scaled entropy'~$\Sigma$ to~$-\infty$ is an artefact of the approximations
made in deriving~\eqnref{entropydef}.  To make this clear, we present here
a brief derivation of~\eqnref{sigmadef} which concentrates on the steps
\emph{before} Eq.~\eqnref{entropydef} rather than, as in
Sect.~\ref{s:assumptions}, on the physics behind the definition
of~$\Sigma$.

As above, we can write~\eqnref{fng}
\[
	f(\Vec r,\Vec v) = n(\Vec r) G(\Vec v),
\]
where~$n(\Vec r)$ and~$G(\Vec v)$ are position and velocity distribution
functions, respectively.  Since these are independent, the statistical
weight for the distribution~$f(\Vec r,\Vec v)$ -- the number of ways of
realising it -- is simply~$\Omega[f] = \Omega[n(\Vec r)]\times\Omega[G(\Vec
v)]$, so that the entropy of the distribution~$f(\Vec n,\Vec v)$ will be
the \emph{sum} of the entropies in position and velocity spaces separately;
this means that we can immediately ignore the constant contribution
of~$n(\Vec r)$ to the entropy, and instead concentrate on the entropy of
the velocity distribution alone.  In these terms, this is
\be{SigmaG}
	S[G] = k \ln \Omega[G],
\ee
where~$k$ is Boltzmann's constant, and~$\Omega[G]$ is the statistical
weight of the distribution, or the number of ways in which the distribution
can be realised.  Note that, since~$\Omega[G]$ is a positive integer, the
entropy~$S[G]$ has a minimum of zero (which occurs when~$G(\Vec v)$ is a
delta function -- there is only one way of giving all the particles the
same velocity).

First define a dimensionless `velocity'~$\Vec u\equiv \Vec v/v_0$, for some
arbitrary speed~$v_0$.  Imagine dividing the accessible volume of velocity
space into a finite number~$\nu$ of small, but not infinitesimal,
volumes, size~$\du$, at~$\Vec u_i$, so that there will be~$N_i \equiv
G(\Vec u_i)\du$ particles in the volume centred on~$\Vec u_i$.
Consequently, the total number of ways of distributing the~$N$ particles
amongst the cells will be the multinomial
\iffalse
\be{Omega1}
	\Omega  = {N! \choose N_1! \cdots N_\nu!}
			= {N! \over \prod_{i}N_i!},
\ee
or
\be{Omega2}
	\ln\Omega = \ln N! - \sum_i \ln N_i!
\ee
\else
$\Omega = N! /\prod_{i}N_i!$.
\fi
If we take~$\du$ to be large enough that each of the~$N_i$ is large, then
we can approximate~$\ln\Omega$ using Stirling's formula~$\ln N!
\approx N \ln N - N$, to obtain
\be{Omega3}
	\ln\Omega \approx -N \sum_i g(\Vec u_i) \ln g(\Vec u_i) \du.
\ee
where we have used the normalisation
\be{Gnorm}
	\sum_i N_i = \sum_i G(\Vec u_i) \du = N,
\ee
and then used the largeness of~$G(\Vec u_i)\du$ to ignore~$\ln\du$,
compared with~$\ln G(\Vec u_i)$, before substituting $G(\Vec u) \equiv
Ng(\Vec u)$.  If, finally, the scale~$\du$ is small enough that we can take
the distribution function~$g(\Vec u)$ to be constant over it, then we can
approximate this final estimate for the distribution
function~\eqnref{Omega3} by an integral, and write the
entropy~\eqnref{SigmaG} as
\be{Sigmag}
	\Sigma[g] \equiv S[g]/Nk =
	    - \int g(\Vec u) \ln g(\Vec u) \, \d^3 \Vec u,
\ee
where we have replaced the approximation sign by an equality.
Compare~\eqnref{sigmadef}.  Crucially, the validity of the
approximation~\eqnref{Sigmag} depends on the distribution function~$G(\Vec u)$
being such that there can be in fact a scale~$\du$ which simultaneously has
sufficiently large~$N_i$ and sufficiently constant~$g(\Vec u)$.  If this is not
so, we generate the artefacts described in Sects.~\ref{ss:compareBIT}
and~\ref{ss:comparePL}.



\end{document}